\documentclass[manuscript=article]{achemso}
\usepackage[version=3]{mhchem}

\usepackage{graphicx}% Include figure files
\usepackage{dcolumn}% Align table columns on decimal point
\usepackage{bm}
\usepackage{enumitem}
\usepackage{graphicx}
\newlist{steps}{enumerate}{1}
\setlist[steps, 1]{label = Step \arabic*:}
\usepackage{listings}
\usepackage[utf8]{inputenc}
\usepackage{amsfonts}
\usepackage[T1]{fontenc}
\usepackage{mathptmx}
\usepackage{xcolor}
\usepackage[title]{appendix}
\newcommand{\bra}[1]{\ensuremath{\left\langle#1\right|}}
\newcommand{\ket}[1]{\ensuremath{\left|#1\right\rangle}}

\newcommand{\bracket}[2]{\ensuremath{\left\langle #1 \middle| #2 \right\rangle}}
\SectionNumbersOn

\author{Zeyu Zhou}
\author{Yanze Wu}
\author{Xuezhi Bian}
\author{Joseph Eli Subotnik}
\email{subotnik@sas.upenn.edu}
 \affiliation[penn]{Department of Chemistry, University of Pennsylvania, Philadelphia, Pennsylvania 19104, United States}

\title[]
  {Non-adiabatic Dynamics in a Continuous Circularly Polarized Laser Field with Floquet Phase-space Surface Hopping}
\begin{document}

\begin{abstract}
Non-adiabatic chemical reactions involving continuous circularly polarized light (cw CPL) have not attracted as much attention as dynamics in unpolarized/linearly polarized light. 
However, including circularly (in contrast to linearly) polarized light allows one to effectively introduce a complex-valued time-dependent Hamiltonian, which offers a new path for control or exploration through the introduction of Berry forces. 
Here, we investigate several inexpensive semiclassical approaches for modeling such nonadiabatic dynamics in the presence of a time-dependent complex-valued Hamiltonian, beginning with a  straightforward 
instantaneous adiabatic fewest-switches surface hopping (IA-FSSH) approach (where the electronic states depend on position and time), continuing to a standard 
Floquet fewest switches surface hopping (F-FSSH) approach (where the electronic states depend on position and frequency), and ending with an exotic 
Floquet phase-space surface hopping (F-PSSH) approach (where the electronic states depend on position, frequency, and momentum).
Using a set of model systems with time-dependent complex-valued Hamiltonians,  we show that the Floquet phase-space adiabats are the optimal choice of basis as far as accounting for Berry phase effects and delivering accuracy.
Thus, the F-PSSH algorithm sets the stage for modeling nonadiabatic dynamics under strong externally pumped circular polarization in the future.
\end{abstract}

\section{\label{sec:one}Introduction}
Non-adiabatic transitions between electronic states typically arise in two different contexts. First, transitions occur naturally through vibronic interactions when molecules visit regions of configuration space where the Born-Oppenheimer approximation is violated; second, transitions can be induced by photo-excitations when an external incident light is coupled to the transition dipole moment between these electronic states. 
Both processes are very important in the field of photochemistry and spectroscopy,\cite{mai2020molecular, nelson2020non,stock2005classical, levine2007isomerization,penfold2018spin}  and both processes need not occur exclusively (i.e. both can occur at the same time).
One important difference between vibronic couplings and light-induced couplings is that the latter is time-dependent; a typical light source contains a central frequency $\omega$ such that the coupling contains $\cos\omega t$. Experiments have shown that strong monochromatic continuous wave (cw) light can change the landscape  of potential energy surfaces and reaction channels by introducing light-induced states, or Floquet states in several different ways; e.g., there is now experimental evidence of light-induced conical intersections.\cite{moiseyev2008laser, halasz2011conical, halasz2012light, corrales2014control, natan2016observation, csehi2017intrinsic, leclerc2017exotic, csehi2017competition, szidarovszky2018direct, triana2019revealing, kubel2020probing, zanchet2020signature, farag2021polariton, fabri2021signatures}

During the past thirty years, various semiclassical formalisms for studying non-adiabatic phenomena have been demonstrated as effective and reasonably accurate.
More recently, many surface hopping  formalisms \cite{tully1990molecular} have been generalized to incorporate time-dependent radiative couplings for light-induced non-adiabatic processes.\cite{mitric2009laser, mitric2011field,lisinetskaya2011simulation, richter2011sharc, bajo2012mixed, mai2015general, fiedlschuster2016floquet, fiedlschuster2017surface, mai2020molecular, zhou2020nonadiabatic}
One of the most intuitive formalisms is to instantaneously calculate the adiabatic potential energy surfaces, whereby one instantaneously diagonalizes the light-matter Hamiltonian.  As a result, time-derivative coupling matrix elements $\mathbf{T}_{jk}=\bra{\psi_{j}(\mathbf{R}(t), t)}\frac{d}{dt}\ket{\psi_{k}(\mathbf{R}(t), t)}$ contains contributions that arise from both nuclear motion and from the explicit time-dependence of the external field $\Bigl\{\ket{\psi_{k}(\mathbf{R}(t), t)}\Bigr\}$. In some cases, these resulting dynamics can perform well\cite{richter2011sharc, mai2015general, mai2020molecular, zhou2020nonadiabatic}, but the algorithm faces a difficult choice when simulating  energy absorption/emission from the external light; usually, when deciding whether a hop is accepted or frustrated, one compares the bandwidth and the energy differences during the nonadiabatic transitions. Energy conservation as a function of photon number is difficult to implement and so the algorithm can lose accuracy.

Apart from IA-FSSH, another possible generalization to the time-dependent nonadiabatic problem is to apply Floquet theory\cite{hone1997time, grifoni1998driven, fiedlschuster2016floquet, fiedlschuster2017surface, zhou2020nonadiabatic}, which transforms a time-periodic Hamiltonian into a time-independent one with larger dimensions, leading to Floquet fewest switches surface hopping algorithm (F-FSSH). 
In our own experience, we have found that, in the presence of monochromatic light (\textit{and} provided the frequency of the light is not very small), F-FSSH performs better in most cases than IA-FSSH insofar as the algorithm better captures energy absorption/emission, which is manifested as transitions between Floquet states with different Fourier indices; interestingly, when the frequency of light is small,  IA-FSSH performs better, as the Hamiltonian approaches a time-independent form. 
For the most part, one can use these two algorithms to reasonably capture most standard nonadiabatic dynamics in a linearly polarized light field.

Now, if we wish to study dynamics under a circular polarized light field (CPL) with large frequency, and if we adopt an electric dipole Hamiltonian, it is fairly easy to conclude that, in a Floquet representation, the light-matter coupling terms become inherently complex-valued (and this  complex-valued nature cannot be eliminated by any simple gauge transformation).
For instance, for a two-level electronic Hamiltonian subject to an external circularly polarized laser (e.g. $\mathbf{E}(t) = E_{x}\cos\omega t\hat{x} + E_{y}\sin\omega t\hat{y}$), in the bare electronic basis and under electric dipole approximation, the light-matter coupling becomes
\begin{equation}
    \bra{\mu}H(t)\ket{\nu} = \mathbf{D}\cdot\mathbf{E}(t) = \mu_{x}E_{x}\cos\omega t + \mu_{y}E_{y}\sin\omega t 
    \label{eq: realtdcoupele}
\end{equation}
Here, $\mathbf{D}$ is the transition dipole moment between electronic states $\ket{\mu}$ and $\ket{\nu}$. 
Note that this coupling term is completely real and time-dependent.
Next, if we apply a Fourier transform, the resulting light-matter couplings in Floquet basis will be of the form 
\begin{equation}
    \bra{m\mu}H(t)\ket{(m\pm\tilde{1})\nu} = \mu_{x}E_{x}/2 \pm i\mu_{y}E_{y}/2
    \label{eq: complexfhdip}
\end{equation}
Here, $m$ is the Fourier index. Clearly, these light-matter couplings between Floquet states with $\pm\tilde{1}$ Fourier indices difference are complex-valued. Note that the transition dipole moment will usually depend on the nuclear configuration and hence, the phases of these coupling terms are not identical for different nuclear configurations.

Unfortunately, the introduction of a complex-valued Hamiltonian renders most of surface hopping schemes inapplicable.  
On the one hand, it is widely acknowledged that during a hopping event between two multidimension potential energy surfaces, to enforce energy conservation, the momentum of the trajectory is rescaled along the direction of derivative coupling $\mathbf{d}$.\cite{jasper2003improved} 
Thus, it is not obvious what direction to choose when the derivative coupling $\mathbf{d}$ is complex-valued. \cite{miao2019extension, miao2020backtracking}
On the other hand, complex-valued derivative coupling $\mathbf{d}$ can alter the nuclear motion by Berry phase effects. 
As is well-known, Berry curvature emerges when the derivative couplings become complex-valued, resulting in an effective magnetic field for the nuclear degrees of freedom.

Over the last few years, several attempts have been made to extend the standard fewest switches surface hopping (FSSH) approach so as to treat a complex-valued Hamiltonian by finding a good rescaling direction and  incorporating Berry phase effects\cite{miao2019extension, miao2020backtracking, bian2021modeling, bian2021meaning, bian2022incorporating, wu2021semiclassical,wu2022phase}.
Within our group, we have developed two such algorithms:
\begin{enumerate}
\item FSSH with ad hoc Berry forces and "$\mathbf{h}$ + $\mathbf{k}$" rescaling direction in ref \citenum{wu2021semiclassical} (FSSH h+k).
The basic premise of FSSH h+k is to take into account the complex nature of the derivative couplings $\mathbf{d}$, which leads to a local effective magnetic field known as the Berry curvature. For this algorithm, the rescaling direction after a hop depends lies in the plane spanned by the derivative of the norm  of the diabatic Hamiltonian elements ($\mathbf{h}$) and the the derivative of the phase $\mathbf{k}$ (see section \ref{subsubsec: FFSSH h+k} for definitions). 

\item Phase-space surface hopping algorithm in ref \citenum{wu2022phase} (PSSH). 
The basic premise is to transform the complex-valued Hamiltonian into a real-valued one \textit{locally} by introducing a phase factor that induces a momentum shift during an electronic transition. 
\end{enumerate}

To date, neither of these algorithms (or any other FSSH algorithm, as far as we are aware) has been successfully extended so as to model the nonadiabatic dynamics of light under a CPL laser field. The goal of this article is to create and benchmark such extensions.

Finally, before concluding this section, we note that CPL is not the only means by which one can introduce complex-valued Hamiltonians (and Berry forces) into nonadiabatic light-matter systems. More generally, Berry forces arise when there is degeneracy of the Hamiltonian, and if one models dynamics without external light -- but with spin degrees of freedom and spin-orbit couplings -- one will also find that a complex-valued Hamiltonian arises.
Moreover, recently there has been speculation that chiral induced spin selectivity (CISS) effects might arise precisely through such coupled nuclear-spin motion. \cite{naaman2012chiral, fransson2020vibrational, hore2020spin} Thus, it is important to emphasize that all of the theory presented below for modeling the dynamics of nuclei and electrons in a circularly polarized light field can be equally applied to modeling the dynamics of nuclei and electrons and spin in a strong linearly polarized light field.

With this background in mind, an outline of this article is as follows: In section \ref{sec:two}, we first review in detail the IA-FSSH and the Floquet-FSSH algorithms which were designed for a system periodically driven by a linearly polarized light. Secondly, we discuss how existing extensions of FSSH to treat complex-valued Hamiltonian can potentially be incorporated into F-FSSH. In section \ref{sec:3}, we  describe the model that we will use to differentiate these different FSSH approaches, and we will offer some visual intuition.
In section \ref{sec:4}, the results of the formalisms discussed above are compared with exact quantum calculations. 
We conclude and discuss several intriguing questions regarding F-PSSH in section \ref{sec:5}.

\section{\label{sec:two} Methods}
\subsection{\label{sec:2.1} Model Hamiltonian and exact solution}
Let us consider a molecule illuminated by continuous wave circularly polarized light (cw-CPL). As discussed in section \ref{sec:one}, the incoming light, in the electric dipole approximation, becomes a real-valued time-periodic coupling  $\hat{V}(t)=\hat{V}(t + T_{0})$ between electronic states with period $T_{0}$ in the original electronic basis (see eq \ref{eq: realtdcoupele}). As shown in eq \ref{eq: complexfhdip}, in the Floquet basis, the coupling becomes time-independent and complex-valued.

As discussed above, if spin degrees of freedom are present, the vibronic coupling can also be complex-valued -- due to spin-orbit coupling or under the influence of an external static magnetic field. Hence, for the most general two level system, let us consider a Hamiltonian of the form:
\begin{eqnarray}
\hat{H}_{\text{tot}}(t)= \hat{\mathbb{T}}_{\mathbf{R}} + \hat{H}_{el}^{0} + \hat{V}(t) = \hat{\mathbb{T}}_{\mathbf{R}} + \hat{H}_{el}(t)
\label{eq:one}
\end{eqnarray}
Here, $\hat{\mathbb{T}}_{\mathbf{R}}$ is the kinetic operator and $\hat{H}_{el}^{0}$ is the time-independent part of the complex-valued electronic Hamiltonian $\hat{H}_{el}$ which we write as 

\begin{align}
\hat{H}_{el}(t) &=  \left[
\begin{array}{c c}
H^{el}_{00}(\mathbf{R}) & H^{el}_{01}(\mathbf{R}, t)\\ 
H^{el}_{10}(\mathbf{R}, t)& H^{el}_{11}(\mathbf{R}) \\ 
\end{array} \right]
\label{eq: timedependentelectronichamiltonian}
\end{align}
Here, the couplings between the two electronic states contain two contribution: a time-independent vibronic coupling term and a time-periodic light-induced coupling term:
\begin{align}
    H^{el}_{10}(\mathbf{R}, t) &= D_{a}(\mathbf{R})\exp(i\phi_{a}(\mathbf{R})) + D_{b}(\mathbf{R})\exp(i\phi_{b}(\mathbf{R}))\cos\omega t
    \label{eq: tdelecham}
\end{align}
Here, $D_{a}$ and $D_{b}$ are the effective  time-independent and time-dependent coupling strengths, and $\phi_{a}(\mathbf{R})$ and $\phi_{b}(\mathbf{R})$ are the phases of the couplings.

The exact solution of such a time-dependent nonadiabatic  problem can be obtained by propagating the Schr\"{o}dinger equation on a grid using short time steps and exponentiating $e^{-i\hat{H}_{tot}(t)dt}$ at each time step, where  $\hat{H}_{tot}(t)$ is the time-dependent total Hamiltonian. The goal of this article is to assess cheap, {\em semiclassical} approaches to such propagation.

\subsection{Four semiclassical methods for time-dependent coupled nuclear-electronic dynamics }
\subsubsection{Instantaneous Adiabatic Fewest Switches Surface Hopping\label{sec: 2.2}}
Let us begin by reviewing the simplest extension to original FSSH for model problems with time-dependent Hamiltonian. As proposed by Gonzalez and Marquetand,\cite{richter2011sharc, mai2015general, mai2018nonadiabatic, mai2020molecular} the basic premise is to use the instantaneous adiabatic potential energy surfaces (that are explicitly time-dependent) to replace the adiabatic potential energy surfaces (that are only parameterized by nuclear configuration $\mathbf{R}$). The nuclear degrees of freedom are evolved by Newton's equations of motion
\begin{align}
    \dot{\mathbf{R}} &= \frac{\mathbf{P}}{M}
    \label{eq: newtoneomr}
    \\
    \dot{\mathbf{P}} &= -\frac{\nabla_{\mathbf{R}}E_{\lambda}(\mathbf{R}, t)}{M}.
    \label{eq: newtoneom}
\end{align}
Here, $E_{\lambda}$ is the instantaneous, active adiabatic state of the electronic Hamiltonian eq \ref{eq: timedependentelectronichamiltonian}.
The electronic degrees of freedom are evolved by the electronic time-dependent Schr\"{o}dinger equations.
\begin{align}
\dot{c}_{j} = -\frac{i}{\hbar}E_{j}(\mathbf{R}, t)c_{j} - i\hbar\sum_{k}\mathbf{P}\cdot \mathbf{d}_{jk}c_{k}/M
\label{eq: tdese}
\end{align}
Here, $\mathbf{d}_{jk}$ is the derivative coupling matrix element between instantaneous adiabatic electronic states $\ket{\psi_{j}}$ and $\ket{\psi_{k}}$. In practice, we evaluate the time-derivative matrix instead 
\begin{equation}
    \mathbf{T}_{jk} = \mathbf{P}\cdot \mathbf{d}_{jk}/M = \bracket{\psi_{j}(\mathbf{R}, t)}{\frac{d \psi_{k}(\mathbf{R}, t)}{dt}},
    \label{eq: timederivativematinstant}
\end{equation}
which contains contributions from both the time-dependent part $\bracket{\psi_{j}(\mathbf{R}, t)}{\frac{\partial \psi_{k}(\mathbf{R}, t)}{\partial t}}$ and the nuclear motion part $\bracket{\psi_{j}(\mathbf{R}, t)}{\nabla_{\mathbf{R}}\psi_{k}(\mathbf{R}, t)}\cdot \frac{\mathbf{P}}{M}$.

Similar to FSSH, the hopping probability from active instantaneous adiabatic state $\lambda$ to state $j$ is
\begin{align}
    g_{\lambda j} = \frac{-2\text{Re}(c_{\lambda}c_{j}^{*}T_{k\lambda})dt}{|c_{\lambda}|^2}
    \label{eq: hopprob}
\end{align}
When a hop from state $\lambda$ to state $j$ occurs, the momentum is only rescaled along the direction of $\mathbf{d}_{\lambda j}$, if the energy difference between state $\lambda$ and state $j$ is not within the bandwidth of the time-dependent driving.

\subsubsection{\label{subsec: FFSSH}Floquet Theory and Floquet Fewest Switches Surface Hopping}
In this subsection, we review Floquet theory and Floquet fewest switches surface hopping (F-FSSH).\cite{ho1983semiclassical, hone1997time, bajo2012mixed, fiedlschuster2016floquet, zhou2020nonadiabatic}
For any problem with real-valued periodic Hamiltonian $\hat{H}(t) = \hat{H}(t+T_{0})$, the time-dependent electronic Schr\"odinger equation is
\begin{eqnarray}
i\hbar\frac{\partial}{\partial t}\ket{\Psi(t)} = \hat{H}(t)\ket{\Psi(t)}
\label{eq:TDSE}
\end{eqnarray}
We define the Floquet Hamiltonian as
\begin{eqnarray}
\hat{H}_{F}(t) \equiv \hat{H}(t) - i\hbar\frac{\partial}{\partial t}
\label{eq:twelve}
\end{eqnarray}
and Floquet diabatic basis as
\begin{align}
    \ket{m\mu} = \exp(im\omega t)\ket{\mu}
    \label{eq: floquetbasis}
\end{align}
Here, $\ket{\mu} = \{\ket{0}, \ket{1}\}$ belongs to a set of orthonormal properly chosen diabatic electronic basis. $m$ is the Fourier basis index, which represents the number of photons dressed by the Floquet state $\ket{m\mu}$. Note that here and below, we will use a superscript tilde to differentiate the ``Floquet photon'' index ($m = \tilde{0}, \pm\tilde{1}, ...$) from the diabatic state index ($\ket{\mu} = \ket{0}, \ket{1}$). 

In the basis $\left\{ \ket{m\mu} \right\}$, the elements of the Floquet Hamiltonian $\hat{H}_{F}(t)$ become time-independent:
\begin{align}
[\hat{\cal H}_{F}]_{(n\nu)(m\mu)} &= \frac{1}{T_{0}}\int_{0}^{T_{0}}dt\bra{\nu}\hat{H}_{F}\ket{m\mu}\exp[-in\omega t]
\nonumber \\
&=\frac{1}{T_{0}}\int_{0}^{T_{0}}dt\bra{\nu}\hat{H}(t)\ket{\mu}\exp[-i(n-m)\omega t] + \delta_{\mu\nu}\delta_{m n}n\hbar\omega
\label{eq:FloquetHam}
\end{align}
Hence, another possible way to propagate exact dynamics is to project the initial wavefunction onto the Floquet states with $m = \tilde{0}$ and evolve the system with the propagator for the time-independent Floquet Hamiltonian. For the complex-valued model Hamiltonians in this paper, the explicit matrix form is written in Appendix \ref{apdx: diabFloquethammatform}.

Now, let us briefly review F-FSSH algorithm.\cite{fiedlschuster2016floquet, zhou2020nonadiabatic} The nuclear degrees of freedom $\mathbf{R}, \mathbf{P}$ are evolved by Newton's equations of motion (see eq \ref{eq: newtoneomr} and \ref{eq: newtoneom}) along the active Floquet adiabatic potential energy surface $E^{F}_{\lambda}(\mathbf{R})$ (eigenvalues of the Floquet Hamiltonian in eq \ref{eq:FloquetHam}, see also Appendix \ref{apdx: diabFloquethammatform}) of the trajectory. 
Note that the Floquet adiabatic potential energy surfaces do not explicitly depend on time (which is different from IA-FSSH).
Similar to standard FSSH, 
the electronic degrees of freedom follow eq \ref{eq: tdese} except that (i) the propagation follows the Floquet adiabatic quasi-energy $E_{\lambda}^{F}(\mathbf{R})$ (rather than instantaneous active adiabatic energy $E_{\lambda}(\mathbf{R}, t)$) and  (ii) the relevant time-derivative matrix element  $\mathbf{T}_{jk}^{F} = \bracket{\psi_{j}^{F}}{\frac{d\psi_{k}^{F}}{dt}}$ is between Floquet adiabatic states (rather than $\mathbf{T}_{jk}$ in eq \ref{eq: timederivativematinstant}).

Lastly, the hopping probability from active Floquet state $\lambda$ to state $k$ is given by the analogue of eq \ref{eq: hopprob}. That being said, there is the question of how to rescale momenta after a hopping event because  $\mathbf{d}_{\lambda j}$ is complex-valued.
To that end, consider a general two-level system of the form:
\begin{align}
    \hat{H}_{el} &=  V\left[
    \begin{array}{c c}
    -\cos\theta & e^{i\phi}\sin\theta\\ 
    e^{-i\phi}\sin\theta& \cos\theta 
    \label{eq: elecham}
\end{array} \right]
\end{align}
For this Hamiltonian, the derivative couplings lie in the vector space spanned by the two directions $\mathbf{h}$ and $\mathbf{k}$
\begin{align}
    \mathbf{h} &= \nabla_{\mathbf{R}}\theta
    \label{eq: hdir}
\\
    \mathbf{k} &= \nabla_{\mathbf{R}}\phi - \frac{\nabla_{\mathbf{R}}\theta(\nabla_{\mathbf{R}}\theta\cdot\nabla_{\mathbf{R}}\phi)}{|\nabla_{\mathbf{R}}\theta|^{2}}
    \label{eq: kdir}
\end{align}
We will follow the convention in ref \citenum{wu2021semiclassical, wu2022phase} and rescale momenta in the direction $\mathbf{h} = \nabla_{R}\theta$.

Finally, at the end of the calculation, there is always the question of how to calculate electronic observables, e.g. the population on a  given electronic state.  For standard FSSH, there is no unique means of calculating such an observable, but from both a theoretical and practical perspective, a density matrix approach usually performs best.\cite{subotnik2016understanding}
Below, we will avoid such nuances and calculated electronic populations only in the asymptotic limit where the diabats and adiabats are equal, and where there is no light-matter coupling. In such a case, the final electronic population can be evaluated (to a good approximation) by summing up the populations of all Floquet states that correspond to a given electronic state:\cite{footnote1}
\begin{align}
    \text{Prob}_{\mu}^{\text{pop}} = \frac{\sum_{m}N_{m\mu}^{\text{traj}}}{N^{\text{traj}}_{tot}}
    \label{eq: finalprob}
\end{align}

\subsubsection{F-FSSH algorithm with Berry force (or ``FSSH with h+k rescaling'') \label{subsubsec: FFSSH h+k}}
The third algorithm that we have tested aims to improve the F-FSSH algorithm by explicitly including  Berry forces along dynamics on one surface \cite{wu2021semiclassical} and making sure that momentum rescaling yield the correct asymptotic values. 
To motivate such an approach and explain how the algorithm  works in practice, let us treat these two effects separately:

{\em Berry Forces}: Recall that Berry forces are effective magnetic fields that emerge with complex-valued Hamiltonians.
The explicit form of the Berry force is:
\begin{align}
    F_{\lambda}^{B} = \frac{2\hbar}{M}\sum_{j}\text{Im}(\mathbf{P}_{\lambda}\cdot\mathbf{d}_{j\lambda})\mathbf{d}_{\lambda j}.
    \label{eq: FB}
\end{align}
As discussed in ref \citenum{wu2021semiclassical}, 
one can calculate the Berry force (eq \ref{eq: FB}) and apply it along  a given FSSH trajectory.

{\em Momentum Rescaling}: For a complex-valued Hamiltonian, as compared to a real-valued one, another relevant difference is that the derivative coupling matrix $\mathbf{d}_{jk}$ becomes complex-valued. Hence, the rescaling direction is no longer well-defined. To that end, we will follow the ansatz in ref \citenum{wu2021semiclassical}. First, even though we have many Floquet states, we calculate the rescaling direction only during a possible hopping event between the active Floquet state and the target Floquet state during a trajectory.
Second, as discussed in ref \citenum{wu2021semiclassical}, when the trajectory has sufficient kinetic energy to hop, we add a component of the momentum along the $\mathbf{k}$ (eq \ref{eq: kdir}) direction, and then we  rescale the momentum (for energy conservation) along $\mathbf{h}$ (eq \ref{eq: hdir}):
\begin{align}
    \mathbf{P}_{k} = \mathbf{P}_{\lambda} + (|\bracket{\zeta_{0}}{k}|^2 - |\bracket{\zeta_{0}}{\lambda}|^2)\hbar\mathbf{k} + \alpha\mathbf{h}
\end{align}
Here, $\zeta_{0}$ is the first diabatic state in eq \ref{eq: elecham}, $\ket{k}$ is the adiabat that the trajectory is hopping to, $\alpha$ is the prefactor determined by energy conservation.
If there is no real solution to $\alpha$, that is, the energy is not sufficient to supplement the momentum change, we use a test momentum $\mathbf{P}^{\text{test}}$ to see if the kinetic energy is consumed because of the Berry force (eq \ref{eq: FB})
\begin{align}
    \mathbf{P}^{\text{test}} = \mathbf{P}_{\lambda} - (|\bracket{\zeta_{0}}{\lambda}|^2 - |\bracket{\zeta_{0}}{\zeta}|^2)\hbar\mathbf{k} + \beta\mathbf{h}
\end{align}
Here, $\zeta$ is the initial diabat.
If there exist a real solution to $\beta$, we set
\begin{align}
    \mathbf{P}_{k} = \mathbf{P}_{\lambda} - \frac{\mathbf{P}\cdot\mathbf{h}}{|\mathbf{h}|^2}\mathbf{h} + \gamma \hbar \mathbf{k}.
\end{align}
Again, $\gamma$ is determined by energy conservation.

The algorithm above is complicated and was developed empirically out of necessity in order to match a set of data.  The intuition is that one ensures the correct asymptotic momentum along $\mathbf{k}$ by depleting the momentum along $\mathbf{h}$.
That being said, if the energy does not allow for a real-valued $\beta$, the hop is frustrated/ (See section \ref{sec:3} for information about velocity reversal.)
Finally, at the end of the calculation, we evaluate the final electronic population in the same fashion as in eq \ref{eq: finalprob}.

\subsubsection{Floquet Phase-Space Surface Hopping Algorithm\label{subsec: fpssh}}
In this subsection, we review our fourth candidate algorithm, a combination of Phase-Space surface hopping (PSSH) together with a Floquet FSSH formalism. \cite{wu2022phase}

Let us begin by reviewing the PSSH algorithm for two electronic states and consider the problem in the form of eq \ref{eq: elecham}.  We will perform a \textit{local} gauge transformation where we introduce a phase on each electronic state (but without rotating the states explicitly):\cite{footnote2}
\begin{align}
    \left[\begin{array}{c}
    \ket{\zeta_{0}'}\\
    \ket{\zeta_{1}'}
    \end{array}\right]
    &= U\left[\begin{array}{c}
    \ket{\zeta_{0}}\\
    \ket{\zeta_{1}}
    \end{array}\right]= \left[
    \begin{array}{c c}
    1 & 0 \\ 
    0 & \exp(-i\phi)
\end{array} \right]
\left[\begin{array}{c}
    \ket{\zeta_{0}}\\
    \ket{\zeta_{1}}
    \end{array}\right]
    \label{eq: gaugetransform1}
\end{align}
The electronic Hamiltonian (eq \ref{eq: elecham}) then becomes real-valued:
\begin{align}
    \hat{H}_{el} &=  V\left[
    \begin{array}{c c}
    -\cos\theta &\sin\theta\\ 
    \sin\theta& \cos\theta
    \label{eq: realelecham}
\end{array} \right]
\end{align}
and the kinetic energy part of the total Hamiltonian becomes
\begin{align}
    \hat{\mathbb{T}}_{\mathbf{R}} = \frac{(\hat{\mathbf{P}} - i\hbar\hat{\mathbf{D}})^{2}}{2M}
\end{align}
Here,  $\hat{\mathbf{D}} = \frac{ U^{\dagger}\hat{P}U - \hat{P}}{i\hbar} =  -i\nabla_{\mathbf{R}}\phi$\ket{\zeta_{1}'}\bra{\zeta_{1}'} where  $\zeta_{1}'$ is the basis after the gauge transform  in eq \ref{eq: gaugetransform1}. The basic idea of PSSH is that one diagonalizes the Hamiltonian $H(R,P) = \hat H_{el} + \mathbb{T}_{\bf R}$ that depends on both position and momentum, and one then moves along the corresponding eigenstates.

Now, let us consider the problem with time-dependent couplings (in eq \ref{eq: tdelecham}) and discuss how to construct the corresponding F-PSSH equations of motion. Once the Hamiltonian (in eq \ref{eq: timedependentelectronichamiltonian}) is turned into a time-independent one as in eq \ref{eq:FloquetHam}, we construct a basis set with dimensionality $N_{elec}N_{Fourier}$, where $N_{elec} = 2$ corresponds to the two electronic states indices and $N_{Fourier}$ is the truncated number of Fourier indices at which the calculation converges (in our calculation $N_{Fourier} = 9$ as $m=\tilde{0}, \pm \tilde{1}, \dots, \pm \tilde{4}$). We seek a local gauge transformation similar to eq \ref{eq: gaugetransform1} and a new basis:
\begin{align}
    \Bigl\{\dots; \ket{\xi^{'(-\tilde{1}) 0}}; \ket{\xi^{'(-\tilde{1}) 1}}; \ket{\xi^{'\tilde{0}0}}; \ket{\xi^{'\tilde{0}1}}; \ket{\xi^{'\tilde{1}0}}; \ket{\xi^{'\tilde{1}1}}; \dots\Bigl\}
    \label{eq: momentumdependentbasis}
\end{align}
where the Floquet Hamiltonian will be strictly real-valued. Here, the $'$ labels the states after the local gauge transformation.

Unfortunately, for this model Hamiltonian (and presumably most Hamiltonians), there is no local gauge transformation under which the Floquet Hamiltonian is strictly real-valued. One possible approximation then is to focus on the most important couplings (i.e. those the couplings mixing populated states that are relevant during a surface hopping calculation) and render those couplings real-valued (or as real-valued as possible). 
Framed mathematically, we seek 
diagonal matrices $\hat{\cal U}_{F}$ and $\hat{\cal D}_{F}$ with the same dimension ($N_{elec}N_{Fourier}$) as the time-independent Floquet Hamiltonian that make the original Floquet Hamiltonian as close to real-valued as possible. 

To construct such matrices explicitly,  let us return to the original Floquet diabatic basis, where the Floquet Hamiltonian is of the form (same as eq \ref{eq:FloquetHam}):
\begin{align}
[\hat{\cal H}_{F}]_{(n\nu)(m\mu)}
&=\frac{1}{T_{0}}\int_{0}^{T_{0}}dt\bra{\nu}\hat{H}_{el}(t)\ket{\mu}\exp[-i(n-m)\omega t] + \delta_{\mu\nu}\delta_{m n}n\hbar\omega
\label{eq:Floquetdiabat2}
\end{align}
with some complex-valued matrix elements arising potentially from both vibronic couplings and light-matter couplings. Eq \ref{eq:FloquetHam} is written out in matrix form in Appendix \ref{apdx: diabFloquethammatform}. Let us suppose (without loss of generality) that the initial state corresponds to $\ket{\tilde{0}0}$.  To make the Hamiltonian as real-valued as possible, we will conjugate this Hamiltonian by a diagonal matrix:
\begin{align}
    \text{diag}(\hat{\cal U}_{F})_{\text{initial state}=\ket{\tilde{0}0}} &= [\dots, e^{i\phi_{a}+\phi_{b}}, e^{i\phi_{b}}, 1, e^{-i\phi_{a}}, e^{-i\phi_{a} - i\phi_{b}}, e^{-i\phi_{b}}, \dots]
    \\
    \text{diag}(\hat{\cal D}_{F})_{\text{initial state}=\ket{\tilde{0}0}} &= [\dots; -\mathbf{w}_{a}-\mathbf{w}_{b}; -\mathbf{w}_{b};0 ; \mathbf{w}_{a}; \mathbf{w}_{a}+\mathbf{w}_{b}; \mathbf{w}_{b};  \dots]
    \label{eq: caldf}
\end{align}
Here,  $\mathbf{w}_{a} = -i\nabla_{\mathbf{R}}\phi_{a}$ and 
$\mathbf{w}_{b} = -i\nabla_{\mathbf{R}}\phi_{b}$
arise from the phase factors in the vibronic coupling in eq \ref{eq: tdelecham}. 
Note that, if the initial state were different (e.g.  $\ket{\tilde{0}1}$), a similar local gauge transformation \Big($(\hat{\cal U}_{F})_{\text{initial state}=\ket{\tilde{0}1}}$\Big) could also be defined. For more details about and an explicit representation of eqs \ref{eq:Floquetdiabat2} - \ref{eq: caldf}, see Appendix \ref{apdx: pfhamnuance}.

At this point, let us assume the Floquet phase-space Hamiltonian is close to real-valued but  depends on both nuclear coordinates $\mathbf{R}$ and momenta $\mathbf{P}$,
\begin{align}
[\hat{\cal H}_{F}]^{'}_{(n\nu)(m\mu)}
&=\frac{1}{T_{0}}\int_{0}^{T_{0}}dt\bra{\nu}\hat{H}_{el}(t)\ket{\mu}\exp[-i(n-m)\omega t] + \delta_{\mu\nu}\delta_{m n}n\hbar\omega - \frac{\hbar^2\hat{\cal D}_{F}^2}{2m} - \frac{i\hbar\hat{P}\cdot\hat{\cal D}_{F}}{m}
\label{eq:Floquetsuperdiabat}
\end{align}
F-PSSH then follows the same procedures as one would expect when combining F-FSSH and PSSH. In a basis of boosted (momentum-dependent) Floquet diabatic states (from eq \ref{eq:Floquetsuperdiabat}), one diagonalizes the momentum-dependent Hamiltonian
and obtains momentum-dependent Floquet adiabatic states, or Floquet phase-space adiabats. All subsequent dynamics move along these phase-space adiabats.  In section \ref{sec:3} below, we will present both the model problem and visualize the difference between Floquet phase-space adiabats and the original Floquet adiabats.

\section{\label{sec:3}Simulation details}
In this paper, we will focus on nonadiabtic models with two nuclear dimensions, one light-induced avoided crossing and one vibronic coupling induced avoided crossing. All parameters are in arbitrary units.
One last small note is now in order. In all of the four algorithms discussed above, if a frustrated hop is encountered, the trajectory is reversed along the rescaling direction $\mathbf{h}$ if $(\mathbf{P}\cdot\mathbf{h})(\nabla_{\mathbf{R}}E^{F}_{j}\cdot\mathbf{h}) > 0$. \cite{jasper2003improved, jain2015surface}

\subsection{Model problems\label{sec:3.1}}
We choose the following diabatic states
\begin{align}
H^{el}_{00}(\mathbf{R}, t) &=A\tanh(Bx)
\\
H^{el}_{11}(\mathbf{R}, t) &=-A(\tanh(Bx) + C)
\end{align}
Here, $A = 0.1$, $B = 0.35$, $C = -0.9$. 
 as they cross near $x=1.5$ and the energy difference between them matches external driving $\hbar\omega = 0.18$ near $x = -1.5$.
For better convergence, the time-dependent couplings (eq \ref{eq: tdelecham}) need to be localized near each type of crossings. Thus, we choose $D_{a}(\mathbf{R})$ and $D_{b}(\mathbf{R})$ to be gaussian functions centered around the two avoided crossings respectively
\begin{align}
    D_{a}(\mathbf{R}) = E \exp(-(x-1.5)^2/2)
    \\
    D_{b}(\mathbf{R}) = E \exp(-(x+1.5)^2/2).
\end{align}
In this paper, $E = 0.02$. We pick simple phase factors as follows: 
\begin{align}
    \phi_{a} &= W_{a}y,\ W_{a} = 6
    \\
    \phi_{b} &= W_{b}y,\ W_{b} = \pm 5
    \label{eq: phasefactors}
\end{align}
These phase factors yield two reasonably strong regions of effective magnetic fields.
Note our notation: while lowercase $\mathbf{w}_{a/b}$ (defined below eq \ref{eq: caldf}) represents the gradient of the phase $\phi_{a/b}$, here we study a problem where both $\mathbf{w}_{a}$ and $\mathbf{w}_{b}$ depend on only one nuclear coordinate, $y$, and so $\phi_{a/b}$ has a gradient along only one direction $y$; above and below, we have denoted this number as $W_{a/b}$.

\begin{figure*}
\includegraphics[width=1.0\textwidth]{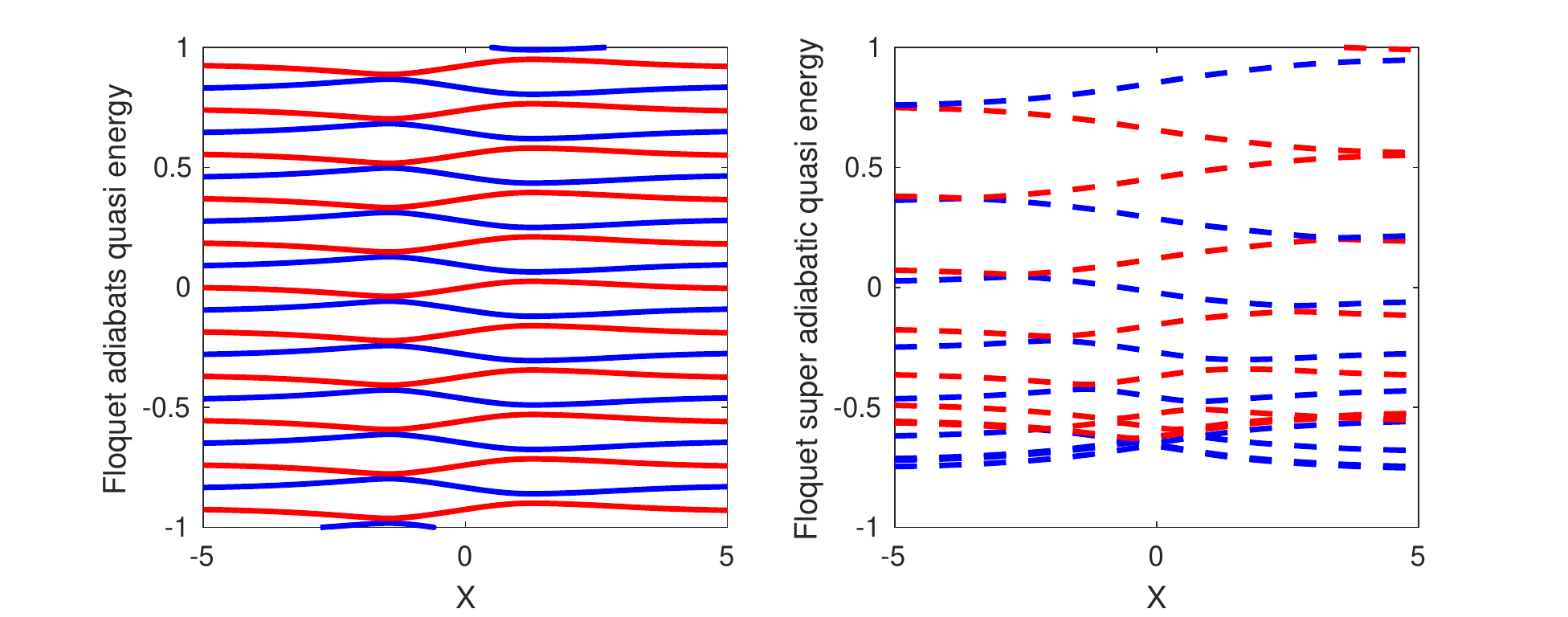} 
\caption{\label{fig: floquetadsuperad} Floquet adiabatic potential quasi-energy surfaces and Floquet phase-space adiabatic potential quasi-energy surfaces for our model. The original F-FSSH and F-FSSH with Berry force algorithms run trajectories along the Floquet adiabatic quasi-energy surfaces; here, the surfaces belonging to the same electronic state are parallel and have equal quasi-energy spacing. The Floquet phase-space adiabatic potential quasi-energy surfaces are parallel but are not equally-spaced (because of the $\hat{\cal D}_{F}^2$ terms in eq \ref{eq:Floquetsuperdiabat}, resulting in $W_a^2$ and $W_b^2$ terms, see also Appendix \ref{apdx: pfhamnuance}). Note that the states drawn on the right are only for normal incidence ($P_{y}(t = 0) = 0$). For oblique incidence, the resulting potential quasi-energy surfaces will also depend on $P_{y}(t = 0)$.}
\end{figure*}

The relevant potential energy surfaces are presented in Figure \ref{fig: floquetadsuperad}. The standard F-FSSH algorithm and F-FSSH with Berry force algorithms evolve trajectories along the standard Floquet states. These Floquet states are parallel and equally spaced. In comparison, for the F-PSSH algorithm, the trajectories run along Floquet phase-space adiabatic states. These states are also parallel, but are not equally spaced, as shown by the dashed lines.

\subsection{Initial Conditions}
For the exact, F-FSSH, F-FSSH with Berry force and F-PSSH calculations, we choose $x(t=0) = -6.0$ which is far enough from $x=-1.5$ such that the initial diabats and adiabats have a  one-to-one correspondence. For all surface hopping algorithms, the initial coordinates and momenta are sampled from the Wigner distribution of the two-dimensional Gaussian wavepacket.

For our two-dimensional model, the effective magnetic fields are parallel to $y$ axis. Thus, we will discuss two possible incident angles, normal incidence ($y(t=0)=0$, $P_{y}(t=0) = 0$) and oblique incidence at $y(t=0) = x_{0}$ and $P_{y}(t=0) = P_{x}(t=0) = P_{init}$ with respect to the $y$ axis in Section \ref{sec:4}.

\subsection{Common techniques}
Before presenting our results, here we will list a few useful tricks and techniques that we used so as to simulate our calculations reliably and efficiently.
\begin{itemize}
    \item Separation of classical and quantum time steps as in ref \citenum{jain2016efficient}. We used a nuclear time step of 0.5 au and an electronic time step is chosen dynamically. 
    
    \item Matrix logarithm to calculate the time-derivative matrix instead of derivative couplings.\cite{hammes1994proton, fabiano2008implementation, barbatti2010non, plasser2012surface, fernandez2012identification, nelson2013artifacts,  wang2014simple, meek2014evaluation, jain2016efficient, dell2017importance, lee2019solving}
    This approach allows for a larger time step without sacrificing any accuracy at all.
    
    \item Parallel transport by maximal phase alignments as in ref \citenum{zhou2019robust}.  Surface hopping within Floquet theory is especially sensitive to the phases of the adiabatic states, and these phases must be chosen accurately.
\end{itemize}

\section{\label{sec:4}RESULTS}
In this section, we first present the transmission and reflection probabilities obtained by exact calculations, F-FSSH, IA-FSSH and F-PSSH. Second, we compare results between F-PSSH and F-FSSH with Berry force.
\subsection{F-PSSH vs standard F-FSSH and IA-FSSH algorithms}
\subsubsection{Normal incidence}
\begin{figure*}
\includegraphics[width=1.0\textwidth]{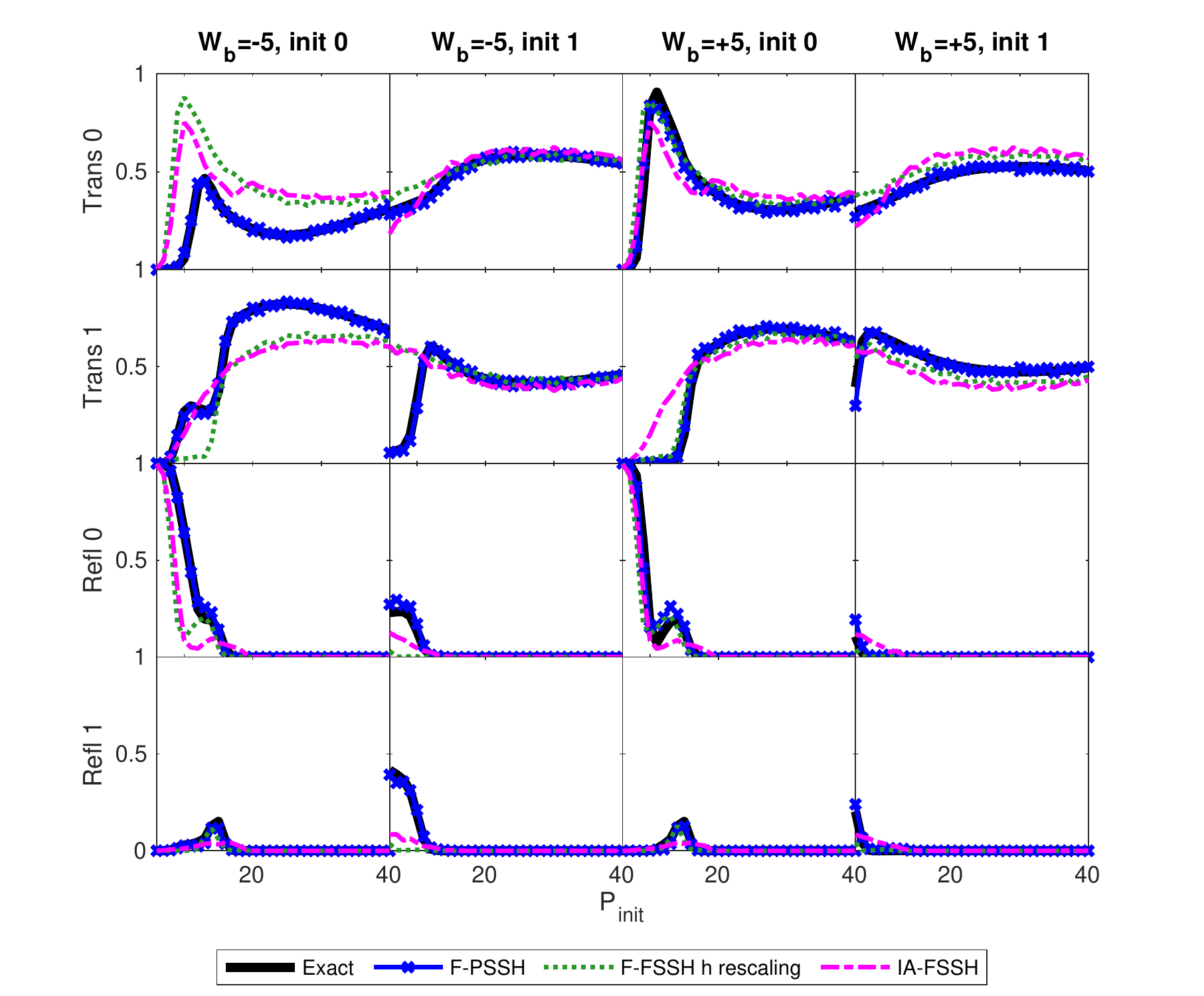}
\caption{\label{fig: d002normal} Transmission and reflection probabilities at normal incidence ($P_{y}(t=0) = 0$) for different choices of initial diabatic electronic states (init $0$ or init $1$) and $W_{b}=\pm 5$. As shown in all subfigures, the F-PSSH algorithm (blue line) yields the most accurate results as compared against the exact results (black line). Note that for the cases with opposite time-dependent phase factors ($W_{b} = -5$ vs $W_{b} = +5$), the standard F-FSSH scheme (green dotted line) predicts the same results (as does IA-FSSH). However, as shown by the exact results, these opposite $W_{b}$ cases are not identical (because of the crossing at $x=1.5$) and admit different scattering probabilities. Thus, this model problem highlights why one requires a nonadiabatic algorithm that can correctly treat complex-valued Hamiltonians.}
\end{figure*}

In Fig \ref{fig: d002normal}, we present the transmission and reflection probabilities on electronic states $0$ and $1$ for different $W_{b}$ and initial electronic states. Several observations can be made.

First, in all of the subfigures, the F-PSSH results (blue line with crosses) agree perfectly with the exact results (black solid line). 
Second, standard F-FSSH (green line, labeled as F-FSSH h rescaling) cannot capture the correct transmission probabilities for the case $W_{b} = -5$ but does give a reasonably good answer for the case  $W_{b} = +5$. 
Interestingly (and incorrectly),  F-FSSH predicts almost the same results for $W_{b} = -5$ vs $W_{b} = +5$. In truth, however, exact scattering results are different for these cases  because the presence of a crossing around $x=1.5$ with complex-valued vibronic couplings ($W_{a} = +6$) breaks any symmetry in $W$.  (If we were to set $W_{a} = 0$, then we would recover the same results for $\pm W_{b}$).
Third, the results obtained by IA-FSSH (magenta line) algorithm are almost the same as standard F-FSSH. In the low momentum regime with $W_{b}=+5$, IA-FSSH is less accurate than F-FSSH. As with F-FSSH, IA-FSSH  incorrectly predicts identical results for $W_{b} = -5$ vs $W_{b} = +5$, which suggests that the algorithm will have similar difficulties with complex-valued Hamiltonians.

\subsubsection{Oblique incidence}
\begin{figure*}
\includegraphics[width=1.0\textwidth]{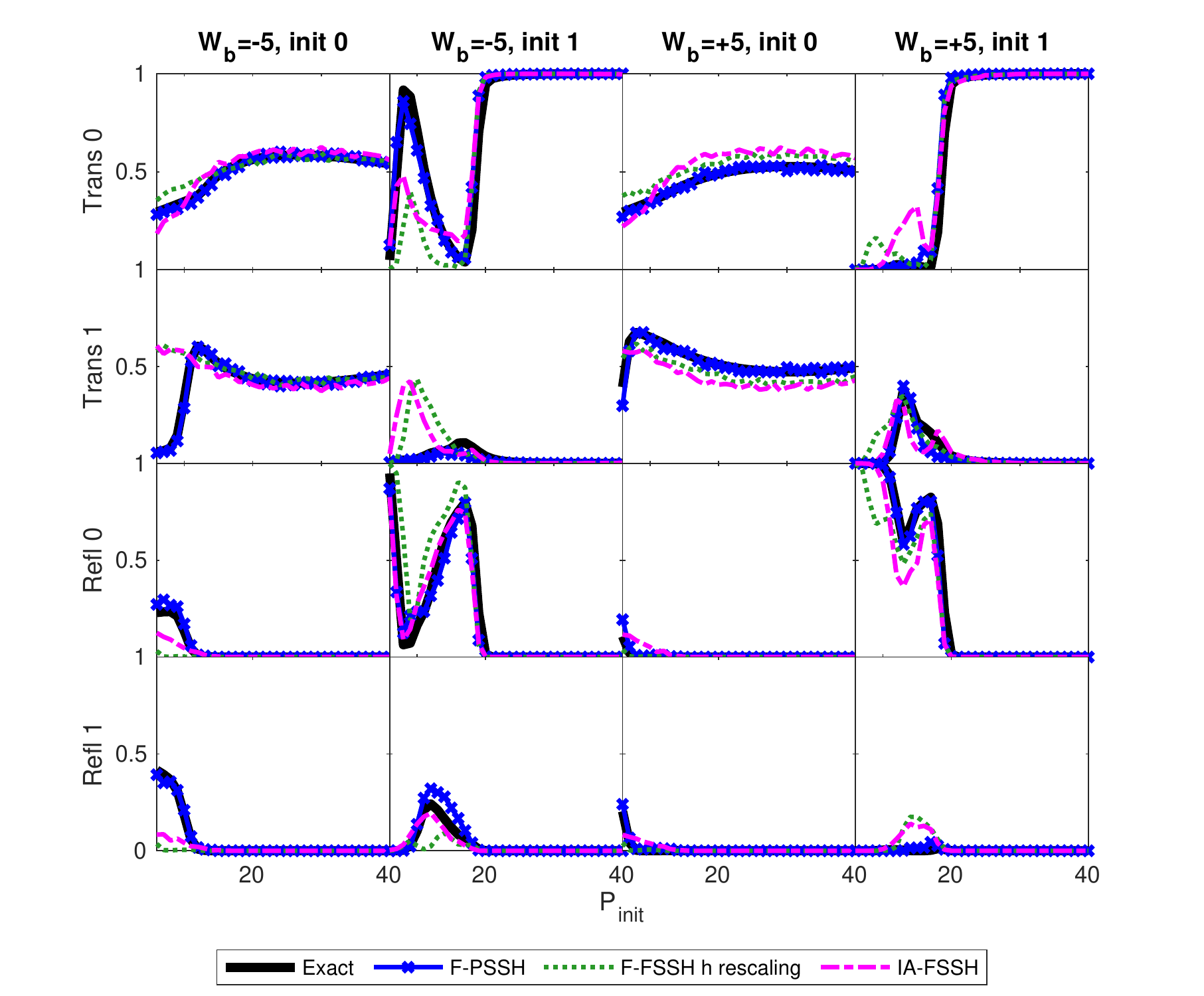}
\caption{\label{fig: D002obliq} Transmission and reflection probabilities for the case of oblique incidence ($P_{y}(t=0) = P_{x}(t=0)$) for different light-induced phase factors ($W_{b}=\pm 5$) and different initial electronic states (init $0$ or init $1$). The F-PSSH results overlap with the exact results in all figures. As in Figure \ref{fig: d002normal}, standard F-FSSH (with $h$-rescaling) and IA-FSSH cannot capture the differences between $W_{b} = -5$ vs $W_{b} = +5$, and the results are thus inaccurate. Moreover, these two algorithms yield incorrect results in the low momentum regime $P_{init} < 20\ a.u.$. Overall, F-PSSH algorithm is clearly the most reliable algorithm.}
\end{figure*}

Next, in Fig \ref{fig: D002obliq}, we present the transmission and reflection probabilities for the case of oblique incidence ($P_{y}(t=0) = P_{x}(t=0) = P_{init}$). This regime is a more difficult test of a surface hopping algorithm. Similar to Figure \ref{fig: d002normal}, F-PSSH does accurately recover the exact results in almost all scenarios. In contrast, standard F-FSSH and IA-FSSH cannot give satisfying results except in the high momentum regime ($P_{init} > 20 \ a.u.$), where effectively $P_{y}(t)$ is not sensitive to the relatively small oscillations in phase ($W_{a/b} << P_{y}$) that each trajectory experiences when moving along the initial electronic state. Note that standard F-FSSH and IA-FSSH algorithms still yield approximately the same results for opposite $W_{b}$ ($W_{b}=-5$ vs $W_{b} = +5$), as they lack the ability to capture the effects of complex-valued couplings. Overall, F-PSSH algorithm outperforms IA-FSSH and standard F-FSSH.
\subsection{F-PSSH vs F-FSSH with Berry force}
\begin{figure*}
\includegraphics[width=1.0\textwidth]{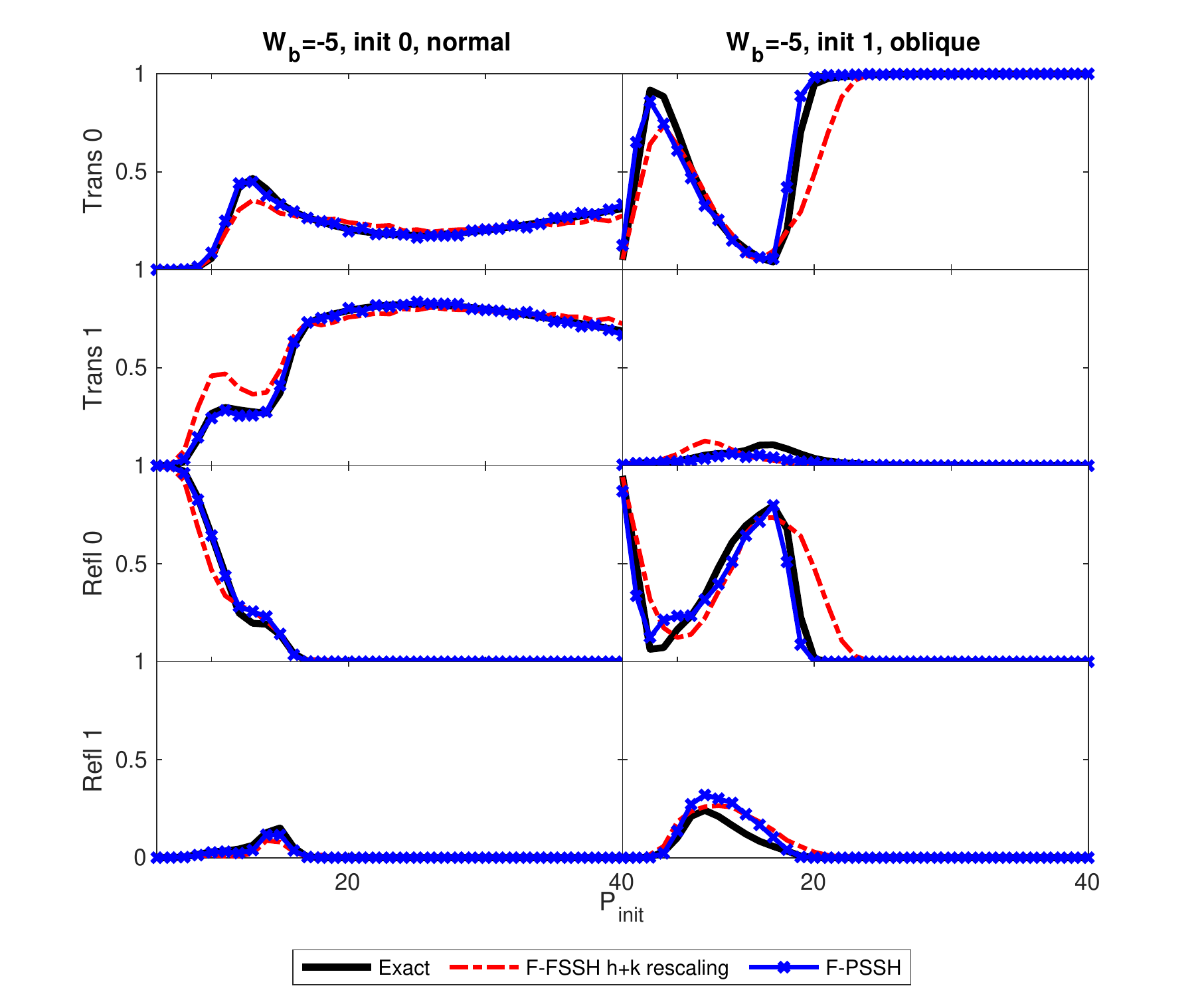}
\caption{\label{fig: D002obliqhk} Transmission and reflection probabilities for $W = -5$ with normal ($\mathbf{P}(t=0) = (P_{init}, 0)$) and oblique incidence ($\mathbf{P}(t=0) = (P_{init}, P_{init})$). In this figure, we compare F-FSSH with $\mathbf{h} + \mathbf{k}$ rescaling vs exact and F-PSSH results. Both of these algorithms include Berry force in some fashion and both are expected to perform well.  In all figures here, we find that  the F-FSSH with $\mathbf{h} + \mathbf{k}$ rescaling algorithm can capture the correct probabilities at high momentum ($P_{init} > 20$) and is reasonably good at low momentum as well; nevertheless, the algorithm is outperformed by F-PSSH in the low momentum regime where predicting the final $\mathbf{P}$ is crucial for capturing scattering probabilities. Nevertheless, if we were to plot F-FSSH $\mathbf{h} + \mathbf{k}$ rescaling  data within Fig \ref{fig: d002normal} and Fig \ref{fig: D002obliq}, we would find that the algorithm  predicts results just as accurately as does F-PSSH.}
\end{figure*}

So far, we have demonstrated that F-PSSH results agree well with the exact results whereas the F-FSSH and IA-FSSH algorithms do not--presumably because the latter two algorithms do not include an Berry force effects. To better probe the value of a phase-space surface hopping code, in this subsection, we will present a detailed comparison of the F-PSSH algorithm with the F-FSSH with Berry force algorithm -- in other words, we compare two semiclassical algorithms that both do account for Berry force. 

In most scenarios treated within Fig \ref{fig: d002normal} and Fig \ref{fig: D002obliq} (and in most other scenarios we tested), F-PSSH and F-FSSH with h+k rescaling (and Berry force) yield equally accurate results. However, for the two specific scenarios  shown in Fig \ref{fig: D002obliqhk}, the F-FSSH with Berry force algorithm (red dashed line, labeled as F-FSSH h+k rescaling) is able to give accurate predictions only for high initial momentum ($P_{init} > 20 \ a.u.$)\cite{wu2021semiclassical}. 
In the low initial momentum regime, however, the F-FSSH with Berry force algorithm allows for an erroneous $P_{y}$ during the simulation run, and this error eventually leads to major deviation from the exact results. While we are convinced that the F-PSSH algorithm is likely the most accurate approach in general, F-FSSH with Berry force may be a satisfactory ansatz in many cases.

\section{\label{sec:5}Conclusions and Discussions}
In summary, we have considered four different surface hopping algorithms:
\begin{itemize}
    \item instantanously adiabatic (IA)-FSSH
    \item Floquet-FSSH
    \item Floquet-FSSH with Ad Hoc Berry Forces
    \item Floquet-PSSH
\end{itemize}
We have benchmarked these algorithm as far as treating a time-dependent and complex-valued Hamiltonian, as relevant to coupled nuclear-electronic problems in a circularly polarized light field). 

For our model problems, both IA-FSSH and the standard F-FSSH algorithms completely fail to capture the resulting Berry phase effects acting on the nuclear motion and transmission/reflection probabilities which arise from
the complex-valued nature of the Hamiltonian.
Between the remaining two algorithms which do include Berry force (at least to some extent), the algorithm with an {\em ad hoc} Berry force and rescaling direction ($\mathbf{h}+\mathbf{k}$) can predict accurate results in the high momentum regime; the Floquet phase-space surface hopping algorithm can recover the exact results almost always both qualitatively and quantitatively.

Interestingly, as a side note, the Hamiltonian in eqs \ref{eq: timedependentelectronichamiltonian}, \ref{eq: tdelecham} and \ref{eq: phasefactors} is spatially periodic in $Y$ (with period  $2\pi$, or angular frequency $\Omega_{Y} = 1$). Thus, just as in Bloch theory, it would appear natural to construct an electronic basis (with a phase parameterized by the nuclear position in the $Y$-direction) of the form
\begin{align}
    \ket{\Xi^{pm\mu}} = \exp(ip\Omega_{Y}Y ) \exp(im\omega_{T} t)\ket{\mu}
\end{align} 
where $\omega_{T}$ is the temporal frequency of external periodic driving and $p$ is a Fourier index. 
Exploring such a basis (and relating this basis back to the basis in eqs \ref{eq: floquetbasis} and \ref{eq: momentumdependentbasis}) might be very revealing as far as simulating systems with both time and spatial symmetry.\cite{flaschner2016experimental}

Looking forward, we can find several paths for future exploration with many open questions.
A first question is, as discussed in section \ref{sec:two}, how to properly calculate final electronic populations. Formally, whewn working within a Floquet picture, one should account for non-zero interference terms of the form: \cite{ho1983semiclassical, zhou2020nonadiabatic}
\begin{equation}
\text{Prob}^{\text{interference}}_{\nu} = \sum_{n\neq m}\frac{\sum_{r=1}^{N_{n\nu}^{\text{traj}}}\sum_{s=1}^{N_{m\nu}^{\text{traj}}}\tilde{c}^{r}_{n\nu}(\tilde{c}^{s}_{m\nu})^{*}\exp(i(n-m)\omega t)}{{N^{\text{traj}}_{n\nu}\times N^{\text{traj}}_{m\nu}}}.
\label{eq:FFSSH_interf}
\end{equation}
Thus, in the context of F-PSSH, one may wonder if/how we might be able to properly calculate the interference terms between wavepackets with different momenta. This question should be addressed in  the future.

A second and equally important question relates to the ubiquitous extra decoherence that must be included within all surface hopping algorithms, i.e. the need to account for wave packet separation. Do existing decoherence ansatzes for FSSH needs to be modified when applied to PSSH approaches, and in particular to F-PSSH?\cite{schwartz1994aqueous, bittner1995quantum, schwartz1996quantum, prezhdo1997evaluation, prezhdo1997mean, fang1999improvement, fang1999comparison, volobuev2000continuous, hack2001electronically, horenko2001theoretical, wong2002dissipative, wong2002solvent, horenko2002quantum, jasper2005electronic, bedard2005mean, subotnik2011decoherence, subotnik2011new, landry2012recover} Our intuitions is that, as in standard FSSH, for F-PSSH the coherence of a trajectory moving along one Floquet phase-space adiabat trajectory  should also be damped as wavepackets on different Floquet phase-space adiabat separate after passing through the crossing region. But for any Floquet based scheme, we have an interesting scenario whereby we have two groups of parallel surfaces, with forces $F_0$ and $F_1$, and we know that decoherence should be proportional to $F_0 - F_1$. Thus,  one might wonder: should there be a unique decoherence rate for every pair of states, $\ket{\xi^{'\tilde{a}0}}$ and $\ket{\xi^{'\tilde{b}1}}$ for all possible $a,b$ photon indices? Or should there be just one effective decoherence rate for the system, capturing an effective rate of wavepacket separation between all those wavepackets on Floquet surfaces $a0$ and all those wavepackets on Floquet surfaces $b1$? 
 
Detailed benchmarking needs to be done before drawing any definitive conclusions. 

Third and finally, the exact functional form of phase factors $\phi$ can be obtained only by performing  {\em ab initio} calculations. As far as the implications of a circularly polarized light field and the corresponding Berry forces are concerned, the most essential question is: how strong are the phase oscillation ($\nabla_{\mathbf{R}}\phi$) for realistic systems and are these oscillations localized or delocalized across configuration space.  These questions require further investigation as well.

In the end, although there are many questions left unanswered, the possibility of using circularly polarized light to push forward coupled nuclear-electronic dynamics is a tantalizing prospect with a largely underexplored knob to control the dynamics (the ellipcity of the light). We believe the Floquet PSSH approach proposed here should offer an essential tool going forward to begin such exploration.

\begin{acknowledgement}
This work was supported by the U.S. Department of Energy, Office of Science, Office of Basic Energy Sciences, under Award No. DE-SC0019397
\end{acknowledgement}
\begin{appendices}
\section{Diabatic Floquet Hamiltonian\label{apdx: diabFloquethammatform}}
In this appendix, for visual ease, we write out explicitly the Floquet Hamiltonian in matrix form.  Our basis $\{\ket{m\mu}\}$ is: 
\begin{align}
    \Bigl\{\dots; \ket{(-\tilde{1}) 0}; \ket{(-\tilde{1}) 1}; \ket{\tilde{0}0}; \ket{\tilde{0}1}; \ket{\tilde{1}0}; \ket{\tilde{1}1}; \dots\Bigl\}
    \label{eq: basis}
\end{align}
Here again, $m = \tilde{0}, \pm \tilde{1}, ...$ represents the Floquet photon indices and $\mu = 0, 1$ represents the electronic state indices.
The explict matrix form of eq \ref{eq:FloquetHam} is (see also eq \ref{eq: timedependentelectronichamiltonian} and \ref{eq: tdelecham}):
\begin{eqnarray}
    & \hat{\cal H}_{F} =  \label{eq: tidfloquethammat}\\
    \nonumber
    &\left(\begin{array}{cccccccc}
    \ddots & \vdots &  \vdots & \vdots & \vdots & \vdots & \vdots & \reflectbox{$\ddots$} \\
    \dots & H^{el}_{00}-\hbar\omega & D_{a}e^{i\phi_{a}} & 0 & D_{b}e^{i\phi_{b}}/2 & 0 &0  & \dots  \\ 
    \dots & D_{a}e^{-i\phi_{a}}& H^{el}_{11}-\hbar\omega  & D_{b}e^{i\phi_{b}}/2 & 0 & 0 &0 & \dots \\
    \dots & 0 & D_{b}e^{-i\phi_{b}}/2 & H^{el}_{00} & D_{a}e^{i\phi_{a}} & 0 & D_{b}e^{i\phi_{b}}/2 & \dots \\
    \dots & D_{b}e^{-i\phi_{b}}/2 & 0 & D_{a}e^{-i\phi_{a}}& H^{el}_{11} & D_{b}e^{i\phi_{b}}/2  &0 & \dots \\
    \dots & 0 & 0 & 0 & D_{b}e^{-i\phi_{b}}/2& H^{el}_{00}+\hbar\omega & D_{a}e^{i\phi_{a}}  & \dots \\
    \dots & 0 & 0 & D_{b}e^{-i\phi_{b}}/2 & 0& D_{a}e^{-i\phi_{a}} & H^{el}_{11}+\hbar\omega & \dots \\
    \reflectbox{$\ddots$} & \vdots & \vdots & \vdots & \vdots & \vdots & \vdots & \ddots
    \end{array}
    \right)
\end{eqnarray}
Note that we omit here all dependence on nuclear configuration $\mathbf{R}$. 
\section{Constructing A Boosted Electronic Hamiltonian that is as Real as Possible for a Phase-space Floquet Hamiltonian Approach\label{apdx: pfhamnuance}}
In section \ref{subsec: fpssh}, we argued that the appropriate gauge transformation $\hat{\cal U}_{F}$ must depend on the initial electronic state of a given problem. This scenario differs from original PSSH algorithm in Ref.  \citenum{wu2022phase}, where the initial state was irrelevant. The difference arises now from the fact that for the  Floquet Hamiltonian as explicitly shown in eq \ref{eq: tidfloquethammat}, it is impossible to define a gauge transformations such that the entire  Hamiltonian becomes real-valued. To see this, let us attempt to make the $6\times6$ block in eq \ref{eq: tidfloquethammat} real. 

Let us denote the target $6$ basis functions as
\begin{align}
    &\left[\begin{array}{c}
    \ket{\xi^{'(-\tilde{1}) 0}}\\
    \ket{\xi^{'(-\tilde{1}) 1}}\\
    \ket{\xi^{'\tilde{0}0}}\\
    \ket{\xi^{'\tilde{0}1}} \\ \ket{\xi^{'\tilde{1}0}}\\ \ket{\xi^{'\tilde{1}1}}\\
    \end{array}\right]
    = \hat{\cal U}_{F} \left[\begin{array}{c}
    \ket{(-\tilde{1}) 0}\\
    \ket{(-\tilde{1}) 1}\\
    \ket{\tilde{0}0}\\
    \ket{\tilde{0}1}\\
    \ket{\tilde{1}0}\\
    \ket{\tilde{1}1}
    \end{array}\right]
    = \left[
    \begin{array}{c c c c c c}
    e^{i\eta_{1}} & & & & & \\ 
     & e^{i\eta_{2}} & & & & \\
     & & e^{i\eta_{3}} & & & \\
     & & & e^{i\eta_{4}} & & \\
     & & & & e^{i\eta_{5}} & \\
     & & & & & e^{i\eta_{6}}
\end{array} \right]
\left[\begin{array}{c}
    \ket{(-\tilde{1}) 0}\\
    \ket{(-\tilde{1}) 1}\\
    \ket{\tilde{0}0}\\
    \ket{\tilde{0}1}\\
    \ket{\tilde{1}0}\\
    \ket{\tilde{1}1}
    \end{array}\right]
    \label{eq: gaugetransform2}
\end{align}
In order to reach a real-valued Hamiltonian, the relative phase differences between the basis functions need to satisfy the following overdetermined set of  algebraic equations:
\begin{align}
    &\left[
    \begin{array}{c c c c c c}
    1 & -1 & & & & \\ 
      &  & 1 & -1 & & \\
      & & & & 1 & -1 \\
      & 1 & -1 & & & \\
    1 & & & -1 & & \\
      & & & 1 & -1 & \\
      & & 1 & & & -1
\end{array} \right]
\left[\begin{array}{c}
    \eta_{1}\\
    \eta_{2}\\
    \eta_{3}\\
    \eta_{4}\\
    \eta_{5}\\
    \eta_{6}
    \end{array}\right]
    =\left[\begin{array}{c}
    \phi_{a}\\
    \phi_{a}\\
    \phi_{a}\\
    \phi_{b}\\
    \phi_{b}\\
    \phi_{b}\\
    \phi_{b}
    \end{array}\right]
    \label{eq: gaugetransform3}
\end{align}

The first three rows arise from forcing the complex-valued vibronic couplings to be real-valued,  and the last four rows arise from forcing the complex-valued light-matter couplings to be real-valued. However, there is no solution to these algebraic equations. For instance, if we sum over rows $1, 2, 4$ and compare the sum with row $5$, we find:
\begin{align}
    \eta_{1} - \eta_{2} + \eta_{3} - \eta_{4} + \eta_{2} - \eta_{3} &= \eta_{1} - \eta_{4} = 2\phi_{a} + \phi_{b}
    \\
    \eta_{1} - \eta_{4} &= \phi_{b}
\end{align}
For $\phi_{a} \neq k\pi$, $k\in \mathbb{Z}$, there is no solution to this set of algebraic equations. Similarly, when we sum over rows $2, 3, 6$ and compare the sum with row $7$, we obtain the same type of contradiction. In the end, there simply is no gauge transformation ${\cal U}_{F}$ under which the (infinite-dimensional) Floquet Hamiltonian becomes real-valued.

With this constraint in mind, a practical approach forward is to recognize that only a few Floquet states are actually populated during a typical surface hopping calculation. 
After all, for this same reason, the formally infinite dimensional Floquet Hamiltonian can be safely truncated; for our calculations in Fig. \ref{fig: floquetadsuperad} - \ref{fig: D002obliqhk} , the corresponding Floquet Hamiltonians are $18\times 18$ matrices.
Thus, it makes sense for us to concern ourselves and make real-valued only  those states that directly couple to the initial Floquet state.

For the scenario that the initial state corresponds to $\ket{\tilde{0}0}$, we choose to make the couplings to $\ket{\tilde{0}1}$ and $\ket{\tilde{-1} 1}$ real. If we fix $\eta_{3} = 0$, the corresponding matrix of phases is:
\begin{align}
    \text{diag}(\hat{\cal U}_{F})_{\text{initial state}=\ket{\tilde{0}0}} &= [\dots, e^{i\eta_{1}}, e^{i\phi_{b}}, 1, e^{-i\phi_{a}}, e^{i\eta_{5}}, e^{i\eta_{6}}, \dots]
\end{align}
The choice of $\eta_{1}$, $\eta_{5}$, $\eta_{6}$ is irrelevant to the results. Simply for the sake of concreteness, we will choose the phase $\eta_{1}$ for states $\ket{(-\tilde{1})0}$ so as to make the coupling $\bra{(-\tilde{1})0}\hat{\cal H}_{F} \ket{(-\tilde{1})1}$ real-valued, we will choose the phase $\eta_{5}$ for state $\ket{\tilde{1}0}$ so as to make the coupling $\bra{\tilde{0}1}\hat{\cal H}_{F} \ket{\tilde{1}0}$ real-valued and we choose the phase $\eta_{6}$ for state $\ket{\tilde{1}1}$ so as to  make the coupling $\bra{\tilde{0}0}\hat{\cal H}_{F} \ket{\tilde{1}1}$ to be real-valued. The final result is:
\begin{align}
    \text{diag}(\hat{\cal U}_{F})_{\text{initial state}=\ket{\tilde{0}0}} = [\dots, e^{i\phi_{b}+i\phi_{a}}, e^{i\phi_{b}}, 1, e^{-i\phi_{a}}, e^{-i\phi_{a} - i\phi_{b}}, e^{-i\phi_{b}}, \dots]
\end{align}
For the couplings between the rest of the states with larger Floquet photon indices, we follow the same fashion such that the closest complex-valued couplings to the initial state is transformed to be real-valued.

Naturally, there is a different transformation if are to simulate dynamics with $\ket{\tilde{0}1}$ as the
initial state. Now, if we fix $\eta_{4} = 0$, we obtain
\begin{align}
    \text{diag}(\hat{\cal U}_{F})_{\text{initial state}=\ket{\tilde{0}1}} &= [\dots, e^{i\eta_{1}}, e^{i\eta_{2}},
    e^{i\phi_{a}}, 1, e^{-i\phi_{b}},  e^{i\eta_{6}}, \dots] 
    \nonumber\\
    &= [\dots, e^{i\phi_{b}}, e^{i\phi_{a}+i\phi_{b}},
    e^{i\phi_{a}}, 1, e^{-i\phi_{b}},  e^{-i\phi_{b}-i\phi_{a}}, \dots]
\end{align}
As one would expect, the final results do not depend on $\eta_{1}$, $\eta_{2}$, $\eta_{6}$.

The matrices $\hat{\cal D}_{F}$ for these two scenarios are obviously: 
\begin{align}
    \text{diag}(\hat{\cal D}_{F})_{\text{initial state}=\ket{\tilde{0}0}} &= [\dots; -\mathbf{w}_{a}-\mathbf{w}_{b}; -\mathbf{w}_{b};0 ; \mathbf{w}_{a}; \mathbf{w}_{a}+\mathbf{w}_{b}; \mathbf{w}_{b};  \dots]
    \\
    \text{diag}(\hat{\cal D}_{F})_{\text{initial state}=\ket{\tilde{0}1}}&= [\dots-\mathbf{w}_{b}; -\mathbf{w}_{a}-\mathbf{w}_{b}; -\mathbf{w}_{a};0 ; \mathbf{w}_{b}; \mathbf{w}_{a}+\mathbf{w}_{b}; \dots]
\end{align}

Under these gauge transformations, we obtain two diabatic Floquet Hamiltonians respectively:
\begin{eqnarray}
    & (\hat{\cal H}_{F})_{\text{initial state}=\ket{\tilde{0}0}} = \\
    \nonumber
    &\left(\begin{array}{cccccccc}
    \ddots & \vdots &  \vdots & \vdots & \vdots & \vdots & \vdots & \reflectbox{$\ddots$} \\
    \dots & H^{el}_{00}-\hbar\omega & D_{a} & 0 & D_{b}e^{-2i\phi_{a}}/2 & 0 &0  & \dots  \\ 
    \dots & D_{a}& H^{el}_{11}-\hbar\omega  & D_{b}/2 & 0 & 0 &0 & \dots \\
    \dots & 0 & D_{b}/2 & H^{el}_{00} & D_{a} & 0 & D_{b}/2 & \dots \\
    \dots & D_{b}e^{2i\phi_{a}}/2 & 0 & D_{a}& H^{el}_{11} & D_{b}/2  &0 & \dots \\
    \dots & 0 & 0 & 0 & D_{b}/2& H^{el}_{00}+\hbar\omega & D_{a}e^{2i\phi_{a}}  & \dots \\
    \dots & 0 & 0 & D_{b}/2 & 0& D_{a}e^{-2i\phi_{a}} & H^{el}_{11}+\hbar\omega & \dots \\
    \reflectbox{$\ddots$} & \vdots & \vdots & \vdots & \vdots & \vdots & \vdots & \ddots
    \end{array}
    \right)
\end{eqnarray}

\begin{eqnarray}
    & (\hat{\cal H}_{F})_{\text{initial state}=\ket{\tilde{0}1}} = \\
    \nonumber
    &\left(\begin{array}{cccccccc}
    \ddots & \vdots &  \vdots & \vdots & \vdots & \vdots & \vdots & \reflectbox{$\ddots$} \\
    \dots & H^{el}_{00}-\hbar\omega & D_{a}e^{2i\phi_{a}} & 0 & D_{b}/2 & 0 &0  & \dots  \\ 
    \dots & D_{a}e^{-2i\phi_{a}}& H^{el}_{11}-\hbar\omega  & D_{b}/2 & 0 & 0 &0 & \dots \\
    \dots & 0 & D_{b}/2 & H^{el}_{00} & D_{a} & 0 & D_{b}e^{-2i\phi_{a}}/2 & \dots \\
    \dots & D_{b}/2 & 0 & D_{a}& H^{el}_{11} & D_{b}/2  &0 & \dots \\
    \dots & 0 & 0 & 0 & D_{b}/2& H^{el}_{00}+\hbar\omega & D_{a} & \dots \\
    \dots & 0 & 0 & D_{b}e^{2i\phi_{a}}/2 & 0 & D_{a}& H^{el}_{11}+\hbar\omega & \dots \\
    \reflectbox{$\ddots$} & \vdots & \vdots & \vdots & \vdots & \vdots & \vdots & \ddots
    \end{array}
    \right)
\end{eqnarray}
Note that there is a real-valued 3x3 matrix block inside of each Floquet Hamiltonian.

Lastly, by following eq \ref{eq:Floquetsuperdiabat}, if the initial state is $\ket{\tilde{0}1}$, the final  diabatic phase-space Floquet Hamiltonian is:
\begin{align}
    &[\hat{\cal H}_{F}]^{'}(\mathbf{R}, \mathbf{P}) = \\
    \nonumber
    &\left(\begin{array}{cccccccc}
    \ddots & \vdots &  \vdots & \vdots & \vdots & \vdots & \vdots & \reflectbox{$\ddots$} \\
    \dots & H^{el}_{00}-\hbar\omega & D_{a} & 0 & D_{b}e^{-2i\phi_{a}}/2 & 0 &0  & \dots  \\ 
    \dots & D_{a}& H^{el}_{11}-\hbar\omega  & D_{b}/2 & 0 & 0 &0 & \dots \\
    \dots & 0 & D_{b}/2 & H^{el}_{00} & D_{a} & 0 & D_{b}/2 & \dots \\
    \dots & D_{b}e^{2i\phi_{a}}/2 & 0 & D_{a}& H^{el}_{11} & D_{b}/2  &0 & \dots \\
    \dots & 0 & 0 & 0 & D_{b}/2& H^{el}_{00}+\hbar\omega & D_{a}e^{2i\phi_{a}}  & \dots \\
    \dots & 0 & 0 & D_{b}/2 & 0& D_{a}e^{-2i\phi_{a}} & H^{el}_{11}+\hbar\omega & \dots \\
    \reflectbox{$\ddots$} & \vdots & \vdots & \vdots & \vdots & \vdots & \vdots & \ddots
    \end{array}
    \right)
    \\
    \nonumber
    &-\left(\begin{array}{cccccccc}
    \ddots & \vdots &  \vdots & \vdots & \vdots & \vdots & \vdots & \reflectbox{$\ddots$} \\
    \dots & \frac{\hbar^2(\mathbf{w}_{a} + \mathbf{w}_{b})^2}{2m} & 0 & 0 & 0 & 0 &0  & \dots  \\ 
    \dots & 0 & \frac{\hbar^2 \mathbf{w}_{b}^2}{2m}  & 0 & 0 & 0 &0 & \dots \\
    \dots & 0 & 0 & 0 & 0 & 0 & 0 & \dots \\
    \dots & 0 & 0 & 0 & \frac{\hbar^2 \mathbf{w}_{a}^2}{2m} & 0  &0 & \dots \\
    \dots & 0 & 0 & 0 & 0 & \frac{\hbar^2(\mathbf{w}_{a} + \mathbf{w}_{b})^2}{2m} & 0 & \dots \\
    \dots & 0 & 0 & 0 & 0 & 0 & \frac{\hbar^2 \mathbf{w}_{b}^2}{2m} & \dots \\
    \reflectbox{$\ddots$} & \vdots & \vdots & \vdots & \vdots & \vdots & \vdots & \ddots
    \end{array}
    \right)
    \\\nonumber
    &+\left(\begin{array}{cccccccc}
    \ddots & \vdots &  \vdots & \vdots & \vdots & \vdots & \vdots & \reflectbox{$\ddots$} \\
    \dots & \frac{i\hbar\mathbf{P}\cdot(\mathbf{w}_{a} + \mathbf{w}_{b})}{m} & 0 & 0 & 0 & 0 &0  & \dots  \\ 
    \dots & 0 & \frac{i\hbar\mathbf{P}\cdot\mathbf{w}_{b}}{m}  & 0 & 0 & 0 &0 & \dots \\
    \dots & 0 & 0 & 0 & 0 & 0 & 0 & \dots \\
    \dots & 0 & 0 & 0 & -\frac{ i\hbar\mathbf{P}\cdot\mathbf{w}_{a}}{m}  &0 & 0 & \dots \\
    \dots & 0 & 0 & 0 & 0 & -\frac{i\hbar\mathbf{P}\cdot(\mathbf{w}_{a} + \mathbf{w}_{b})}{m} & 0 & \dots \\
    \dots & 0 & 0 & 0 & 0 & 0 & -\frac{i\hbar\mathbf{P}\cdot\mathbf{w}_{b}}{m} & \dots \\
    \reflectbox{$\ddots$} & \vdots & \vdots & \vdots & \vdots & \vdots & \vdots & \ddots
    \end{array}
    \right)
\end{align}

If the initial state is $\ket{\tilde{0}1}$, then the final diabatic phase-space Floquet Hamiltonian is:
\begin{align}
    &[\hat{\cal H}_{F}]^{'}(\mathbf{R}, \mathbf{P}) = \\
    \nonumber
    &\left(\begin{array}{cccccccc}
    \ddots & \vdots &  \vdots & \vdots & \vdots & \vdots & \vdots & \reflectbox{$\ddots$} \\
    \dots & H^{el}_{00}-\hbar\omega & D_{a}e^{2i\phi_{a}} & 0 & D_{b}/2 & 0 &0  & \dots  \\ 
    \dots & D_{a}e^{-2i\phi_{a}}& H^{el}_{11}-\hbar\omega  & D_{b}/2 & 0 & 0 &0 & \dots \\
    \dots & 0 & D_{b}/2 & H^{el}_{00} & D_{a} & 0 & D_{b}e^{-2i\phi_{a}}/2 & \dots \\
    \dots & D_{b}/2 & 0 & D_{a}& H^{el}_{11} & D_{b}/2  &0 & \dots \\
    \dots & 0 & 0 & 0 & D_{b}/2& H^{el}_{00}+\hbar\omega & D_{a} & \dots \\
    \dots & 0 & 0 & D_{b}e^{2i\phi_{a}}/2 & 0 & D_{a}& H^{el}_{11}+\hbar\omega & \dots \\
    \reflectbox{$\ddots$} & \vdots & \vdots & \vdots & \vdots & \vdots & \vdots & \ddots
    \end{array}
    \right) \\
    \nonumber
    &-\left(\begin{array}{cccccccc}
    \ddots & \vdots &  \vdots & \vdots & \vdots & \vdots & \vdots & \reflectbox{$\ddots$} \\
    \dots & \frac{\hbar^2 \mathbf{w}_{b}^2}{2m} & 0 & 0 & 0 & 0 &0  & \dots  \\ 
    \dots & 0 & \frac{\hbar^2(\mathbf{w}_{a} + \mathbf{w}_{b})^2}{2m}  & 0 & 0 & 0 &0 & \dots \\
    \dots & 0 & 0 & \frac{\hbar^2 \mathbf{w}_{a}^2}{2m} & 0 & 0 & 0 & \dots \\
    \dots & 0 & 0 & 0 & 0 & 0 & 0& \dots \\
    \dots & 0 & 0 & 0 & 0 & \frac{\hbar^2 \mathbf{w}_{b}^2}{2m} & 0 & \dots \\
    \dots & 0 & 0 & 0 & 0 & 0 & \frac{\hbar^2(\mathbf{w}_{a} + \mathbf{w}_{b})^2}{2m} & \dots \\
    \reflectbox{$\ddots$} & \vdots & \vdots & \vdots & \vdots & \vdots & \vdots & \ddots
    \end{array}
    \right)
    \\\nonumber
    &+\left(\begin{array}{c c c c c c c c}
    \ddots & \vdots &  \vdots & \vdots & \vdots & \vdots & \vdots & \reflectbox{$\ddots$} \\
    \dots & \frac{i\hbar\mathbf{P}\cdot\mathbf{w}_{b}}{m} & 0 & 0 & 0 & 0 &0  & \dots  \\ 
    \dots & 0 & \frac{i\hbar\mathbf{P}\cdot(\mathbf{w}_{a} + \mathbf{w}_{b})}{m} & 0 & 0 & 0 &0 & \dots \\
    \dots & 0 & 0 & \frac{ i\hbar\mathbf{P}\cdot\mathbf{w}_{a}}{m} & 0 & 0 & 0 & \dots \\
    \dots & 0 & 0 & 0 & 0 & 0 & 0 & \dots \\
    \dots & 0 & 0 & 0 & 0 & -\frac{i\hbar\mathbf{P}\cdot\mathbf{w}_{b}}{m} & 0 & \dots \\
    \dots & 0 & 0 & 0 & 0 & 0 & -\frac{i\hbar\mathbf{P}\cdot(\mathbf{w}_{a} + \mathbf{w}_{b})}{m} & \dots \\
    \reflectbox{$\ddots$} & \vdots & \vdots & \vdots & \vdots & \vdots & \vdots & \ddots
    \end{array}
    \right)
\end{align}
\end{appendices}
\bibliography{FPSSH}

\begin{mcitethebibliography}{80}
\bibitem[foo()]{footnote2}
Note that the $angle \phi$ depends on the nuclear configuration $\mathbf{R}$
  and there is no diabatic basis in which this complex-valued Hamiltonian
  becomes real-valued Hamiltonian.\relax
\mciteBstWouldAddEndPunctfalse
\mciteSetBstMidEndSepPunct{\mcitedefaultmidpunct}
{}{\mcitedefaultseppunct}\relax
\EndOfBibitem
\end{mcitethebibliography}

\end{document}